\documentclass[12pt]{article}
\usepackage{graphicx}
\newcommand{\widthfig}{0.5\textwidth}
\newfont{\boldit}{cmbxti10 scaled \magstep 3}

\title{ENTROPY, PROBABILITY AND DYNAMICS}
\author{E. G. D. Cohen\\
The Rockefeller University\\
New York, NY 10021}
\date{  }
\begin{document}
\maketitle

\begin{abstract}
Boltzmann's struggle with a derivation of the Second Law of
Thermodynamics is sketched.  So is his first derivation of the
connection between entropy and probability in 1877.  Planck's
derivation and quantum mechanical modifications of Boltzmann's
connection between entropy probability are given next.  Then
Einstein's objections to a purely probabilistic rather than a
dynamical interpretation of entropy are discussed. Finally, the
dynamical basis of the Sinai-Ruelle-Bowen distribution for very
chaotic systems is sketched and appears to be an example of
Einstein's dynamical interpretation of entropy.
\end{abstract}

\noindent {\underline{I. Introduction}}\\

Boltzmann had a strong disposition for mechanics and his first
papers were all devoted to purely mechanical derivations of the
Second Law of Thermodynamics. The most important one was that
based on the 1872 Boltzmann equation, i.e. on the dynamics of
binary collisions, where he says at the end:

{\it{``One has therefore rigorously proved that, whatever the
distribution of the kinetic energy at the initial time might have
been, it will, after a very long time, always necessarily approach
that found by Maxwell''$^{[1]}$.}}

He seems to have overlooked entirely that the Stoszzahl Ansatz,
i.e. the assumption of molecular chaos used in his equation, was a
statistical assumption which had no dynamical basis.

It is therefore, in my opinion, ironic that perhaps his most
famous achievement may well have been the relation of 1877 between
entropy and probability, which was devoid of any dynamical
feature$^{[2]}$.

Whenever later dynamical results were obtained, e.g. by Helmholtz,
when he introduced his monocycles in 1884$^{[3]}$, Boltzmann
immediately jumped at it and extended it to what we now call the
origins of ergodic theory, i.e. a dynamical theory$^{[4]}$.  Also
his summarizing lectures in the two volume ``Lectures on Gas
Theory''$^{[5]}$ mostly discuss dynamical approaches and reference
to the entropy-probability relation can mainly be found on a few
pages of the first book in connection with the statistical
interpretation of his $H$-function (ref.5,I,p.38).

However, in this lecture I do want to concentrate on his work on
entropy and probability first and end with a revival of the
dynamical approach as proposed by Einstein and as later used, in
my view, in the dynamical approach to phase space probabilities in
the Sinai-Ruelle-Bowen (SRB) distribution. The sudden switch which
Boltzmann made from a purely dynamical to a purely probabilistic
approach, might well have been due to the critical attacks of many
of his colleagues on the Stoszzahl Ansatz, as exemplified by
Loschmidt's Reversibility Paradox$^{[6a]}$ and Zermelo's
Recurrence Paradox$^{[6b]}$.\\

\noindent{\underline{II. Boltzmann's original derivation of
{$S\sim {\rm{log}} W$}}\\

This was done in a paper of 1887$^{[2]}$: {\it{``On the relation
between the Second Law of the Mechanical Theory of Heat and
Probability Theory with respect to the laws of thermal
equilibrium.''}} I will sketch first the simplified procedure
Boltzmann follows in Chapter I of this paper.

The crucial statement here is: ``For an ideal gas in thermal
equilibrium the probability of the number of ``complexions'' of
the system is a maximum.''

Boltzmann introduced the notion of ``complexions''
as follows:\\
a) Assume discrete kinetic energy values of each molecule, which
are represented in an arithmetic series: {$$\varepsilon,
2\varepsilon, 3\varepsilon, ..., p \varepsilon,$$} where each
molecule can only have a finite number, $p$, of kinetic energies
$\varepsilon$.\\
b) Before each binary collision the total kinetic energy of the
two colliding molecules is always contained in the above series
and ``by whatever cause'' the same is true after the collision.
He says:

{\it{``There is no real mechanical system [to which this collision
assumption is applicable], but the so-defined problem is
mathematically much easier to deal with and [in addition] it goes
over in the problem we want to solve, when the kinetic energies of
the molecules become continuous and $p \rightarrow \infty$''}}.

Assume that the possible kinetic energies of the $N$ molecules are
distributed in all possible ways at constant total kinetic energy
$E$. Then each such distribution of the total kinetic energy over
the molecules is called a {\underline{complexion}}. What is the
number $P$ of complexions, where {$w_j$} molecules possess a
kinetic energy $j \varepsilon \; (j=1, ...., N)$? This number $P$
[which Boltzmann calls ``the permutability'' or ``thermodynamic
probability''] indicates how many complexions correspond to a
given molecular distribution or state of the system. A
distribution can be represented by writing down first as many
$j$'s as there are molecules with a kinetic energy $j \varepsilon
\; (w_j)$ etc. Obviously $P = N!/\Pi^p_{j=1} w_j!$. The most
probable distribution is that for which the $\{w_j\}$ are such
that $P=\max$ or $\Pi^p_{j=1} w_j!$ or also $\log \Pi^p_{j=1}
w_j!$ a minimum. With the constraints $\sum^p_{j=1} w_j = N$ and
$\sum^p_{j=1} (j \varepsilon) w_j= E$ and  Stirling's
approximation, Boltzmann finds then for the probability that the
kinetic energy of a molecule is $s \varepsilon$:
$$w_s \sim -\exp -s\varepsilon/\bar{\varepsilon}$$
with $\bar{\varepsilon} = \frac{E}{N}$, the average kinetic energy
of a molecule.

Boltzmann adds that: {\it{``in order to translate the above
[derivation] into the mechanical theory of heat, in particular
into the introduction of differentials [when one makes the kinetic
energies of the molecules continuous rather than discrete], needs
still some thought and a not unimportant modification of the
formulae.''}}\\
This was worked out in the following three chapters of this paper.

It ultimately leads then to the results that $\max P$ yields
Maxwell's exponential kinetic energy distribution as well as to
the developments in Chapter V, which I will now discuss.

In Chapter V of this paper Boltzmann discusses the {\it{``Relation
of the Entropy to that quantity, which I have called the
[thermodynamic] probability distribution $(9)$''.}} He first makes
the connection between what he calls the {\underline{degree of
permutability}} $\Omega$ or $\log P$ given by:
$$\Omega = -\int \int f{\bf{(r,v)}} {\rm{log}} f{\bf{(r,v)}}
{\bf{dr dv}}$$ and the entropy. Here $f(\bf{r,v})$ is a continuous
generalization of the discrete $w_j$ used before, giving the
number of molecules at the position $\bf{r}$ with velocity
$\bf{v}$. The maximum of $\Omega$ under the constraints that $N$
and $E$ are given, leads then again to the exponential Maxwell
equilibrium velocity distribution. Noting that the degree of
permutability $\Omega$ differs from the logarithm of the
permutability $P$ only by a constant, one has $\Omega = \log P +$
constant. If the gas was initially not in thermal equilibrium and
approaches equilibrium, $\Omega$ must reach a maximum
$\Omega_{max}$. Boltzmann then computes
{\underline{thermodynamically}} the entropy $S$ for an ideal gas
with $N$ particles and average kinetic energy or temperature $T$.
For reversible processes he finds then:
$$S = \int dQ/T = \Omega = \log P + \: {\rm{constant}}.$$

As Abraham Pais remarks in his classic book ``Subtle is the
Lord....''$^{[7]}$:  {\it{``Boltzmann's qualities as an
outstanding lecturer are not reflected in his scientific papers,
which are sometimes unduly long, occasionally obscure and often
dense. Their main conclusions are sometimes tucked away among
lengthy calculations.''}}

The paper I am discussing is a prime example of this description.
Only in the text towards the end of the paper the following
sentence appears referring to the previously obtained formula $S=
\int dQ/T = \Omega$ when he says:

{\it{``Now it is known that when in a system of bodies only
{\underline{reversible}} changes occur, the total entropy of all
these bodies remains constant.  If, however, also
{\underline{irreversible}} processes occur, then the total entropy
of the system must necessarily grow....''}}

According to the equation $$d \Omega = d \log P = dQ/T =dS$$
 the increase in the sum of the degree of permutability of a system
 $d\Omega$ equals then the increase of its entropy $dS$. He says:

{\it{``Therefore the degree of permutability $\Omega$ is a
quantity, which in the equilibrium state, apart from a constant
factor and an addend, is identical with the entropy $S$.''}}

That is: $S=c_1 \Omega + c_2^\prime = c_1 \log P + c_2$, but this
formula is {\underline{not}} in the paper.

Boltzmann ends this very long [60 pages] paper by remarking that
too little is known both experimentally and theoretically about
liquids and solids, to generalize his relation $S = c_1 \log P +
c_2$ from ideal gases to liquids or solids. He remarks that he
{\it{``has given earlier arguments that it is likely that also for
these states of aggregation, thermal equilibrium will be
determined by a maximum of the quantity $\Omega$, which [also] for
{\underline{such}} systems [would be] identical with the entropy
$S$.''}}\\

\noindent{\underline{III. Planck's derivation of Boltzmann's
$S=c_1\log P +c_2$ }}$^{[8]}$\\

It was really Planck who made the step from Boltzmann's paper to
$S=k \ln W+c$, where $W$ is written instead of $P$. Planck bases
his discussion of the connection between entropy and probability
on the universality of both the Second Law and the laws of
probability, {\it{``so that it is to be expected that the
connection between entropy and probability should be very
close''}}.

Hence he makes the following proposition as the foundation of all
further discussion: {\it{``The entropy of a physical system in a
definite state depends solely on the [thermodynamic] probability
$W [\equiv P]$ of this state.''}}

Without knowing this probability $W$, $S$ can be determined as
follows.\\

\noindent {\underline{Planck's derivation of Boltzmann's $S =
f(W)$}}\\

The probability $W$, for a system consisting of two entirely
independent systems with probabilities $W_1$ and $W_2$,
respectively, is: $W = W_1 W_2$. Then $S_1 = f(W_1)$ and $S_2 =
f(W_2)$. Second Law: $S = S_1 + S_2$ or $f(W_1W_2) = f(W) = f(W_1)
+ f(W_2)$. Differentiating both sides of this equation with
respect to $W_1$ with $W_2$ = constant and then the resulting
equation with respect to $W_2$ at $W_1$ = constant one obtains:
$$\dot{f}(W) + W \ddot{f} (W) = 0$$
with the solution: $f(W) = k \log W$ + constant or $S = k \log W +
c.$ Thus Planck formulated Boltzmann's connection between entropy
$S$ and the permutability $P \sim W$, in its definitive form:
$$S = k \log W + c$$
Planck notices two differences between his formula and
Boltzmann's $S = c_1 \log P + c_2$:\\
a) he replaces Boltzmann's macroscopic expression for $c_1 = R/N$
by a molecular quantity which he called Boltzmann's
constant $k$.\\
b) the additive constant $c_2$ is undetermined as is the case in
the whole of classical thermodynamics. Planck assigns a definite
value to $c_2$, i.e. a definite value to $S$, by using the
``hypothesis of quanta''.

That is, he assumed that in every finite region of phase space the
thermodynamic probability has a finite magnitude limited by [the
existence of] $h$, Planck's constant, and can not be
infinitesimally small like in the classical case. This allows $S$
to be determined free of an arbitrary constant and at the same
time to connect the classical and quantum mechanical values of
$S$, such that $S = 0$ at $T = 0$ (Sackur-Tetrode
formula)$^{[9]}$.

To the best of my knowledge neither Boltzmann's nor Planck's
derivation of $S$ for an ideal gas has ever been generalized to an
interacting gas. Results for such a gas can be obtained from
Gibbs' microcanonical ensemble but that is based ultimately on the
unproven ergodic hypothesis.\\

\noindent{\underline{IV. Einstein's Objection$^{[10]}$}}\\

\noindent 1. From 1905-1920 Einstein repeatedly objected to the
probabilistic derivation of $S = k {\rm{log}} W + c$.\\
2. Basis: the thermodynamic probability $W$ to find the system in
a certain complexion can {\underline{only}} be determined
{\underline{dynamically}} and not be guessed statistically by
assigning ``ad hoc permutabilities''
($P$) to complexions of the system.\\
3. In fact, it is determined by the frequency that a phase space
trajectory visits a given region of phase space due to its
{\underline{dynamics}} in phase space.\\

\noindent {\underline{Quote of 1910$^{[10b]}$}}\\

{\it{``Usually $W$ equals the number of complexions.  In order to
compute $W$ [however] one needs a {\underline{complete}}
(molecular-mechanical) theory of the system.  Therefore it is
dubious that the Boltzmann principle has any meaning
{\underline{without}} a {\underline{complete}}
molecular-mechanical theory or some other theory which describes
the elementary [dynamical] processes [of the system].  In
equilibrium, the expression $S = k {\rm{log}} W + c$, seems
[therefore] devoid of [any] content from a phenomenological point
of view,
without giving in addition such an elementary theory.''}}\\

\noindent {\underline{V. The SRB distribution$^{[11]}$}}\\

Recently a theory has been developed for systems in a
{\underline{\it{non}}}-equilibrium stationary state, which are in
principle, not restricted to be near equilibrium. This theory uses
a probability or measure for ``complexions''of the system in its
phase space following Einstein's dynamical rather than Boltzmann's
(probabilistic) Principle, as Einstein called it. This measure is
usually called the Sinai-Ruelle-Bowen or SRB measure$^{[11]}$,
which will be sketched in a physical way now. For simplicity, I
will use a two dimensional representation.\\

\noindent{\underline{SRB distribution (physically)}}\\

Consider a smooth and very chaotic classical dynamical system
(Anosov-like). This Chaoticity is based on the hyperbolicity of
the points representing the system in its phase space. Each such
point has two manifolds (cf.fig.1):\\
a) an {\underline{unstable}} manifold $(u)$, on which two separate
points near a given phase point exponentially
{\underline{separate}} from each other;\\
b) a {\underline{stable}} manifold $(s)$ on which two separated
points near a given phase point exponentially
{\underline{approach}} each other.
\begin{figure}[!htbp]
\begin{center}
\includegraphics[width=\widthfig]{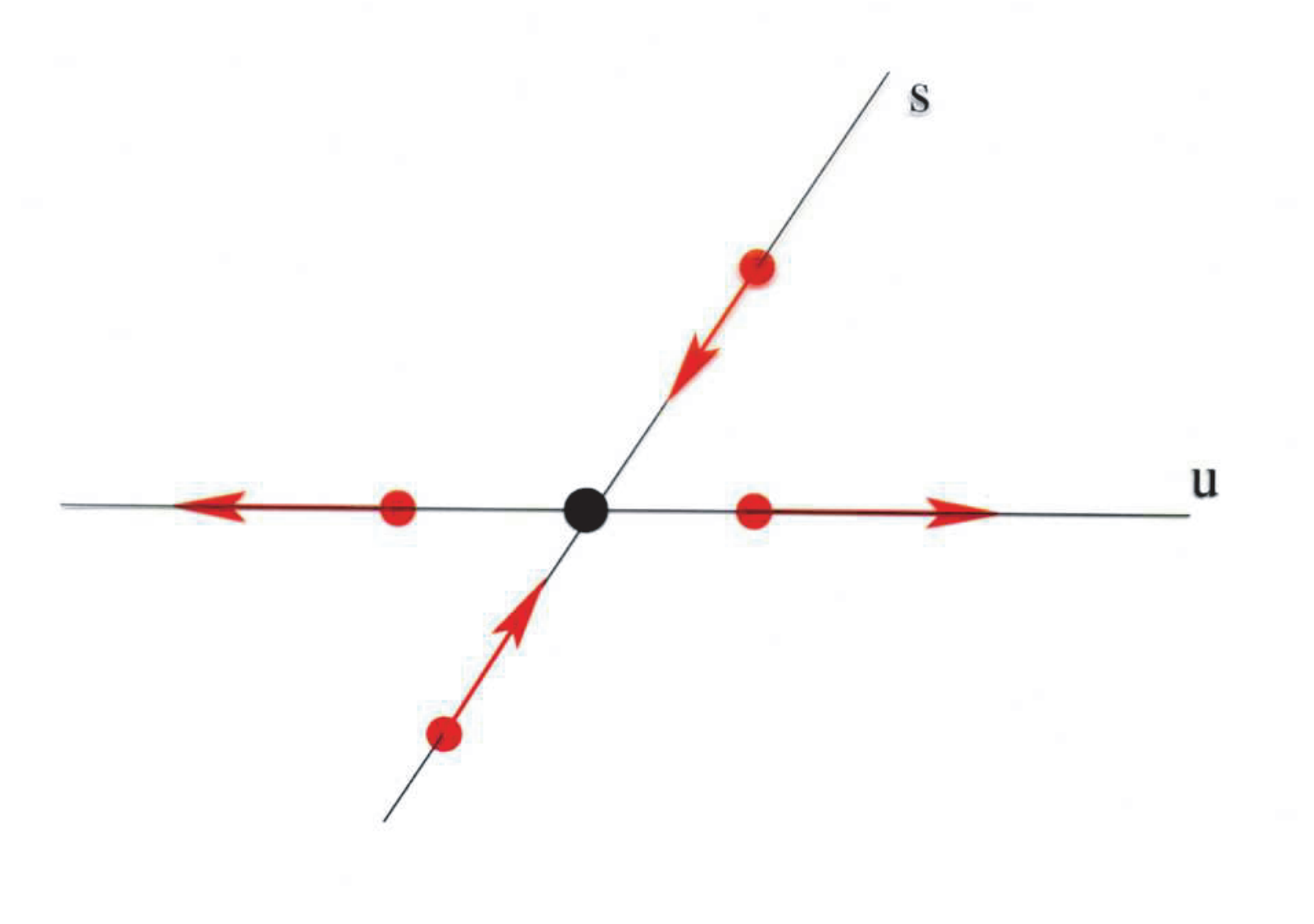}
\caption{Unstable ($u$) and stable ($s$) manifolds associated with a phase point
(thick bullet) in a two dimensional representation.}
\end{center}
\end{figure}
The union of all unstable manifolds of the phase points
representing the system is called the {\underline{unstable
manifold}} in the system's phase space; the union of all stable
manifolds of the phase points representing the system is called
the {\underline{stable manifold}} in the system's phase space. One
now makes a (Markov) partition of the system's phase space into
``parallelograms'' (cells) (cf.fig.2).

The ``horizontal sides'' of the parallelograms form together the
unstable manifold and the ``vertical'' sides of the parallelograms
form together the stable manifold in the phase space. The size of
these parallelograms is determined by a parameter $\cal{T}$, so
that for $\cal{T}$ $\rightarrow 0$ their sizes go to zero.
\begin{figure}[!htbp]
\begin{center}
\includegraphics[width=\widthfig]{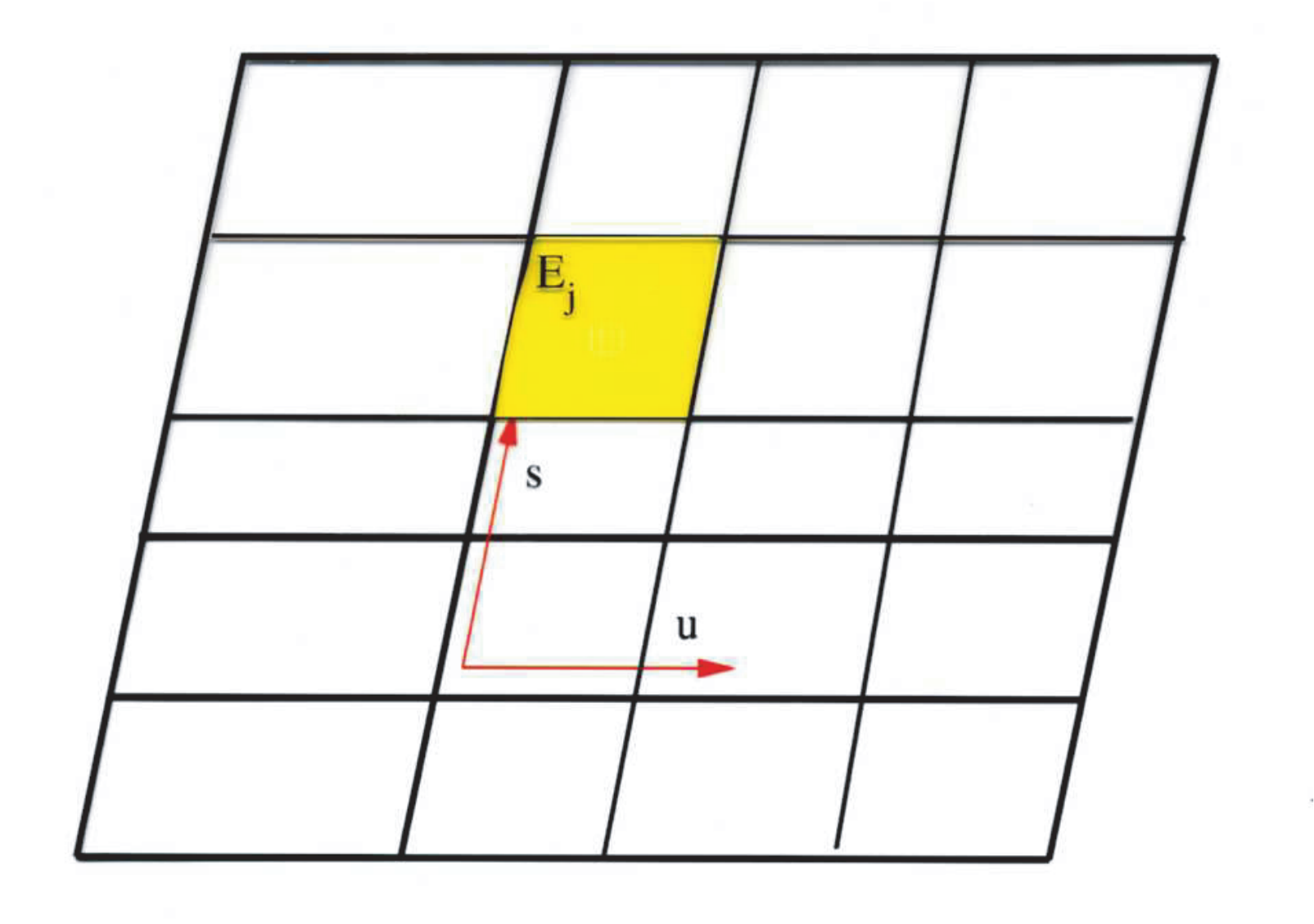}
\caption{Markov partition of a two dimensional phase space with cells $E_j$, formed
by the unstable $(u)$ and stable $(s)$ manifolds of the phase points.}
\end{center}
\end{figure}
Now each cell $E_j$ in phase space is given a weight
$\Lambda^{-1}_{u,\tau} (x_j)$ which is equal to the inverse of the
phase space expansion along the unstable manifold, during a time
$\tau$. This phase space expansion is based on a trajectory moving
during a (discrete) time $\tau$ from $-\tau/2-1$ to $\tau/2$ along
a phase space trajectory through the center $x_j$ of the cell
$E_j$, using the dynamical equations of motion (cf.fig.3).

Considering a small phase space volume $A$ around the initial
point at $-\tau/2-1$ (cf.fig.4), then all points in $A$ will go
via phase space trajectories to corresponding points in the phase
space volume $B$ around the final point at $+\tau/2$. The larger
the phase space volume expansion $\Lambda_{u,\tau}(x_j)$ in the
direction of the unstable manifold $u$ is, i.e. the larger
$L_B/L_A$, the more the phase space trajectories will tend to
avoid (bypass) the point $x_j$.
\begin{figure}[!htbp]
\begin{center}
\includegraphics[width=\widthfig]{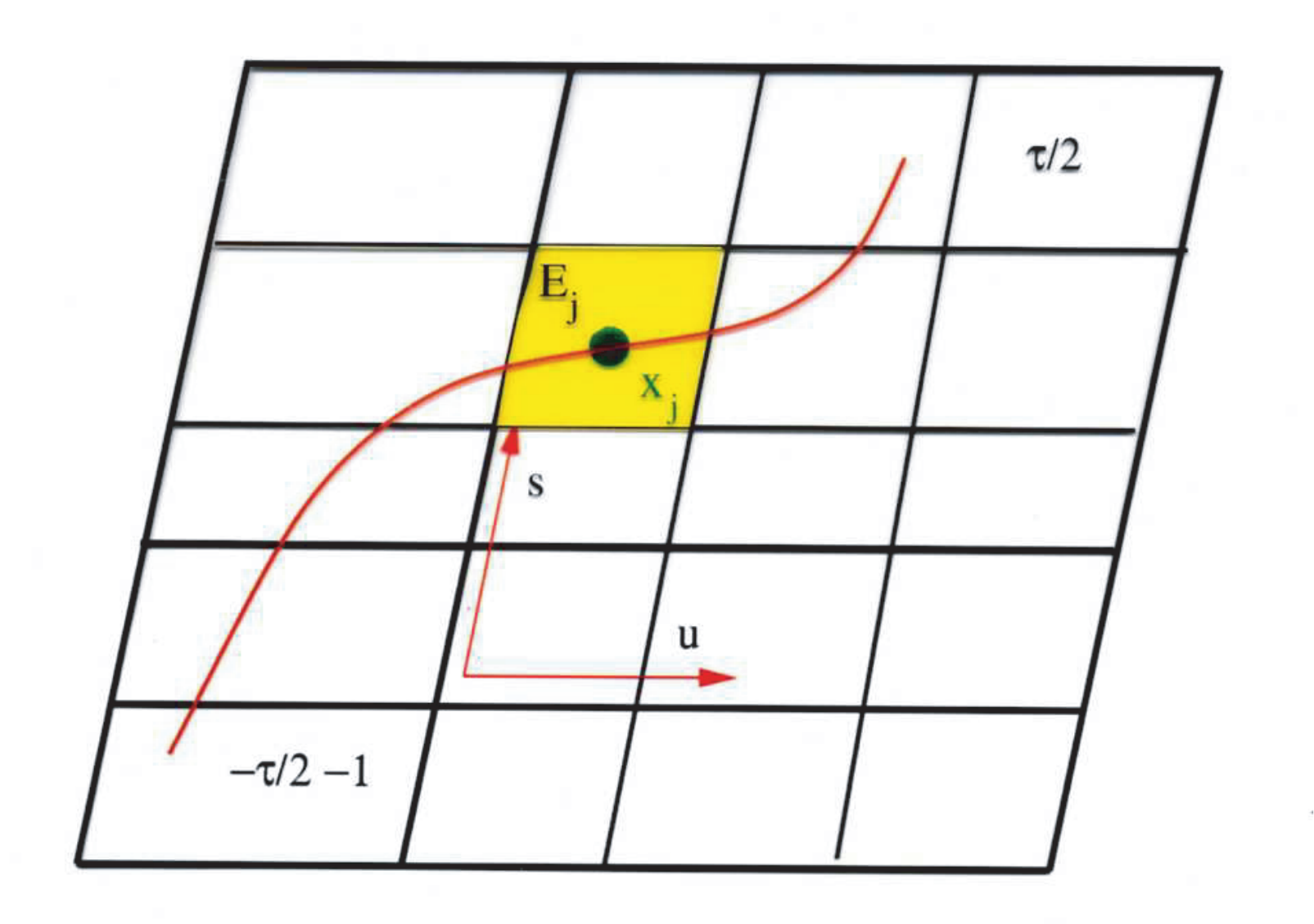}
\caption{Trajectory segment of duration $\tau$ through the center $x_j$ of cell
$E_j$ of a Markov partition of the system in phase space.}
\end{center}
\end{figure}
The inverse of this ratio $\sim L_A/L_B \sim
\Lambda^{-1}_{u,\tau}(x_j)$ will therefore be a measure of the
``eagerness'' or frequency of the phase space trajectories to be
near $x_j$, i.e. that the system will visit the cell $E_j$.
Weighing the Markov partitions in phase space this way, one
obtains in a {\underline{dynamical}} way the probability to find
the system anywhere in phase space. As a consequence, the average
of a smooth function $F(x)$, where $x$ denotes a point in phase
space, is then determined by the SRB measure
$\Lambda^{-1}_{\tau,u} (x_j)$:
$$\int \mu_{SRB}(dx)F(x) = \lim_{{\cal T} \geq \tau/2 \rightarrow
\infty} \; \frac{\sum_j \Lambda^{-1}_{u,\tau} (x_j)F(x_j)}{\sum_j
\Lambda^{-1}_{u,\tau} (x_j)}$$ This appears to me a direct example
of Einstein's proposal of a probability for complexions based on
the system's {\underline{dynamics}} rather than on ad hoc,
although possibly reasonable, probabilistic assumptions as made by
Boltzmann and Planck.
\begin{figure}[!htbp]
\begin{center}
\includegraphics[width=\widthfig]{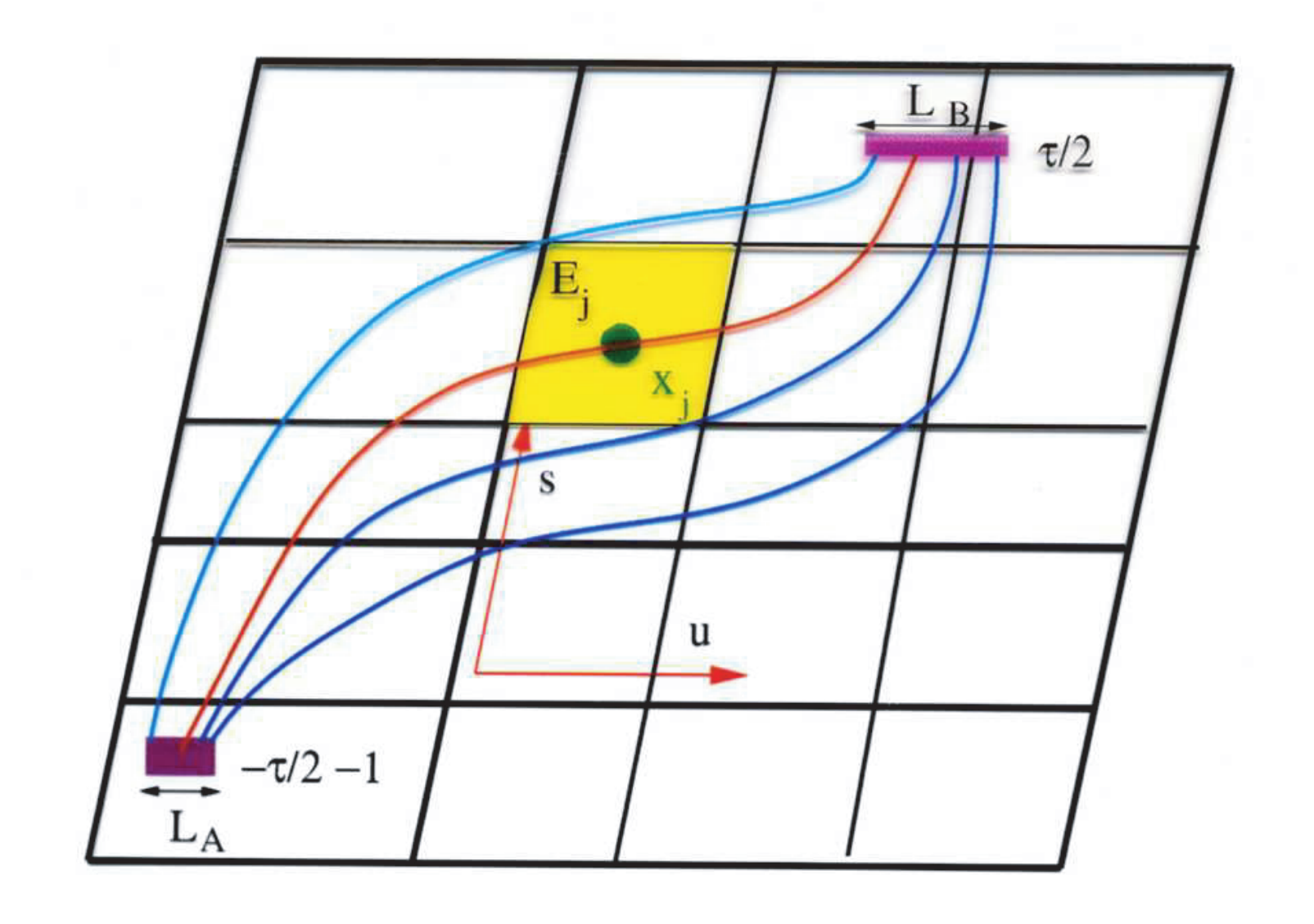}
\caption{Small phase space volume of extension $L_A$ at time $-\tau/2-1$ and of
extension $L_B$, in the direction of the unstable manifold $(u)$ at time $\tau/2$.
$L_B/L_A$ gives the phase space volume increase in the direction of the unstable
manifold $(u)$.}
\end{center}
\end{figure}

\noindent{\underline{VI. Final Remarks}}\\

\noindent 1. In thermal equilibrium the SRB distribution reduces
to Gibbs' microcanonical ensemble. This implies that this ensemble
can also be derived rigorously dynamically along the lines
sketched here, albeit so far only for smooth and sufficiently
chaotic systems. The role of coarse graining phase space into
physically infinitesimal cells, to go from a microscopic to a
macroscopic description, which is used in various ways in all
classical derivations of thermodynamics from statistical mechanics
and also by Boltzmann and Planck, is only used in passing in the
SRB measure, because of the ultimate $\lim_{{\cal T} \geq \tau/2
\longrightarrow \infty}$.

\noindent 2. To actually use Einstein's dynamical method to
determine the probabilities of cells in phase space in practice,
requires a solution of the equations of motion. The only way, so
far, around this fundamental difficulty has been to replace the
effects of dynamics again by suitable probability distributions as
is done e.g. in superstatistics$^{[12-14]}$. This has been applied
successfully to certain distribution functions in turbulent flows
by using e.g. log-normal or $\chi^2$-distributions.

\noindent 3. To the best of my knowledge there is no definition of
entropy beyond the linear (near equilibrium) regime.  This is in
my opinion a fundamental open question, whose solution may be
essential to make real progress into the far from equilibrium
region. Gallavotti and I$^{[15]}$ have conjectured that the
entropy in nonequilibrium stationary states ``far'' from
equilibrium cannot be defined, just like the heat content of a
body in equilibrium cannot be defined, since it depends on how
this heat is used$^{[16]}$. Similarly, entropy
{\underline{differences}} such as entropy production or entropy
transfers could be meaningful also far from equilibrium, just like
heat production or heat transfer are in equilibrium.

\noindent 4. In all fairness to Boltzmann, it should be said that
he also proposed to determine the probability of a state or a
complexion of a system in a region of phase space of this system
in equilibrium, by assuming that this probability would be
proportional to the {\underline{time}} the system spends in it
i.e. involving the dynamics of the system. But for some reason he
did not develop this idea further.

\vspace{.25in}

Finally it would be, in my opinion, a grave mistake to think that
Boltzmann would have been opposed to these new dynamical
developments. As I said before, he had a strong disposition for
dynamics and I think his probabilistic work was presumably
inspired to get rid of his many opponents to his dynamical work
and was just an intermezzo for him, important as it is for us.

In fact, Boltzmann would have been delighted to see that a
dynamical theory of phase space weights was developed and used a
century after his death and that therefore his dynamical
predilection was justified in the end$^{[17]}$.\\

\noindent {\underline{References}}  (W. A. = Wissenschaftliche Abhandlungen)\\

\noindent 1.  Boltzmann, L., {\it{``Weitere Studien \"{u}ber das
W\"{a}rmegleichgewicht unter Gasmolek\"{u}len,''}} Wien. Ber.
\underline{66}, 275-370 (1872); {\it{W. A.}}, Band I, 316-402.\\
2.  Boltzmann, L., {\it{``\"{U}ber die Beziehung zwischen dem
zweiten Hauptsatz der mechanischen W\"{a}rmetheorie und der
Wahrscheinlichkeitsrechnung respektive den S\"{a}tzen \"{u}ber des
W\"{a}rmegleichgewicht''}}, Wien. Ber. \underline{76}, 373-435
(1877); {\it{W. A.}} Band II, 164--223.\\
3. Helmholtz, H. von, Berl. Ber., 6 and 27 March (1884).\\
4. Boltzmann, L., {\it{``\"{U}ber die Eigenschaften monozyklischen
und andere damit verwandter Systeme''}}, Crelles Journal,
(Zeitschr. f. R.U. Angewandte Mathem.) \underline{98}, 68-94 (1884,
1885); {\it{``\"{U}ber die mechanischen Analogien des Zweiten
Hauptsatzes der
Thermodynamik''}}, ibid. \underline{100}, 201-212 (1887).\\
5.  Boltzmann, L., {\it{``Vorlesungen \"{u}ber Gastheorie,''}} I,
II, Barth, Leipzig (1910, 1912).\\
6. Boltzmann, L., (a) {\it{``Bemerkungen \"{u}ber einige Probleme
der mechanischen W\"{a}rmetheorie,''}} Wien. Ber. \underline{74},
62-100 (1877), section II; section II, W. A. Band. III, 116-122;
ibid. (6b) {\it{``Entgegnung auf die W\"{a}rmetheoretischen
Betrachtungen des Hrn.E. Zermelo,''}} Wied. Ann. \underline{57},
778-784 (1896); W. A., Band III, 567-578; ibid., {\it{``Zu Hrn
Zermelo's Abhandlung \"{u}ber die Mechanische Erkl\"{a}rung
irreversibelen Vorg\"{a}nge''}}, Wied. Ann. {\underline{60}},
392-398 (1897); W. A. Band III, 579-586; ibid., {\it{``\"{U}ber
einen mechanischen Satz Poincar\'{e}'s''}}, Wien. Ber.
{\underline{106}},
12-20 (1897); W. A. Band. III, 587-595.\\
7.  Pais, A., {\it{``Subtle is the Lord...''}}, Oxford University
Press (1982) Ch. II.\\
8. Planck, M., {\it{``Theory of Radiation''}}, Dover Publ.
(1951).\\
9.  Tolman, R. C., {\it{``The Principles of Statistical
Mechanics''}}, Dover Publ. (1980) p. 574.\\
10. Einstein, A., (a) {\it{``\"{U}ber einen die Erzeugung und
Verwandlung des Lichtes betreffenden heuristischen
Gesichtspunkt''}}, Ann. d. Physik {\underline{17}}, 132 (1905);
(b) ibid., {\it{``Theorie der Opaleszenz von homogenen
Fl\"{u}ssigkeiten und Fl\"{u}ssig\-keits\-ge\-misch\-en in der N\"{a}he
des kritischen Punktes''}}, Ann. d. Physik. {\underline{33}},
1275-1298 (1910); (c) ibid., Einstein, A., {\it{``Zum
gegenw\"{a}rtigen Stand des
Strahlungs Problems''}}, Physik. Zeitschr. (1909) p. 87.\\
11.  See e.g. Eckmann, J. P. and Ruelle, D., {\it{``Ergodic theory
of Chaos and Strange Attractors''}}, Rev. Mod. Phys.
\underline{57}, 617-656 (1985).\\
12.  Beck, C. and Cohen, E. G. D., {\it{``Superstatistics''}},
Physica A \underline{322}, 267-275 (2003).\\
13.  Cohen, E. G. D., {\it{``Superstatistics''}}, Physica D
\underline{193}, 35-52 (2004).\\
14.  Beck, C, Cohen, E. G. D. and Swinney, H. L., {\it{``From Time
Series to Superstatistics''}},  Phys. Rev. E \underline{72}, 056133 (2005).\\
15. Gallavotti, G. and Cohen, E. G. D., {\it{``Non-equilibrium
Stationary States and Entropy''}}, Phys. Rev. E \underline{69},
035104R (2004).\\
16. Landau, L. D. and Lifshitz, E. M., {\it{``Statistical
Physics''}}, Pergamon Press (1958) p. 43.\\
17. For more historical details, see, e.g. ref. 7 and Cohen, E. G.
D., {\it{``Boltzmann and Statistical Mechanics''}} in:
``Boltzmann's Legacy: 150 Years After His Birth'', Atti Dei
Convegni Lincei \underline{131}, Academia Nazionale Dei Lincei,
Roma, (1997) p. 9-23; also in: {\it{``Dynamics: Models and Kinetic
Methods for Non-equilibrium Many Body Systems''}}, (J. Karkheck,
ed.), Nato Series E, vol. 371, 223-238 (Kluwer, 2006).

\end{document}